\documentclass[12pt, a4paper]{article}
\usepackage{amsfonts, amssymb}
\usepackage{graphicx}
\usepackage{latexsym,amsmath,color,array}

\usepackage{tikz}
\usetikzlibrary{automata, positioning,shapes,snakes}

\usepackage{natbib}
\usepackage{enumerate}
\usepackage{bbding}
\usepackage{amssymb}
\usepackage{amsmath}
\usepackage{graphicx}
\usepackage{amsmath}
\usepackage{bbm}
\usepackage{mathtools}
 \usepackage{color}
 \usepackage{array}
 \usepackage{multirow}
 \usepackage{makecell}
 \usepackage{colortbl}
 \usepackage{subcaption}
 \usepackage{mathrsfs}
 \usepackage{multirow}

\def\1{\mathbb{I}}

\usepackage{amsthm}

\newcounter{example}

\newcounter{remark}

\newcounter{algo}

\newcounter{step}

\begin{document}

\title{An Alternative Formulation of Coxian Phase-type Distributions with Covariates: Application to Emergency Department Length of Stay}
\author{Jean Rizk\footnote{University of Limerick; Jean.Rizk@ul.ie} \hspace{3cm}
Kevin Burke\footnote{University of Limerick; Kevin.Burke@ul.ie} \hspace{3cm}
Cathal Walsh\footnote{University of Limerick; Cathal.Walsh@ul.ie}  }
\date{\today}

\maketitle

\begin{abstract}
In this paper we present a new methodology to model patient transitions and length of stay in the emergency department using a series of conditional Coxian phase-type distributions, with covariates. We reformulate the Coxian models (standard Coxian, Coxian with multiple absorbing states, joint Coxian, and conditional Coxian) to take into account heterogeneity in patient characteristics such as arrival mode, time of admission and age. The approach differs from previous research in that it reduces the computational time, and it allows the inclusion of patient covariate information directly into the model. The model is applied to emergency department data from University Hospital Limerick in Ireland. 

\smallskip

{\bf Keywords.} Coxian phase-type distributions; covariates; length of stay; predictions; emergency department.

\end{abstract}

\qquad

\newpage

\section{Introduction}
The emergency department (ED) is an essential component of the healthcare system as it is the main route of admission to the hospital. The influx of patients into the ED is one of the most challenging problems hospital managers have to deal with.      As the world's population  grows, as well as the proportion of elderly people, who often have complex medical needs,  material resources and staff become overwhelmed, and  the patient flow results in overcrowding. This leads to many serious consequences, such as patients leaving the ED without receiving treatment, ambulances unable to unload their patients, treatment delays, patient elopement, distressed staff, and financial effect \citep{boyle2012, hoot2008}. 

Prolonged length of stay (LoS) is one of the most important causes of overcrowding in the ED. Predicting patient LoS is vital, since it is considered as a proxy for measuring the forthcoming workload and  consumption level of  resources. Therefore, it is necessary  to have reliable measures for modelling and predicting patient LoS in the ED. Patient LoS  data are mainly positively  skewed and heavy tailed. In the last decade, many researcher have shown that phase-type distributions \citep{neuts1975}, particularly Coxian phase-type (CPH) distributions \citep{cox1955}, can accurately  model the patient LoS data in various healthcare systems \citep{marshall2004,vasilakis-adelle2005,shaw2007modelling,marshall2007}. A standard CPH distribution of order $n$, describes duration until absorption in terms of a continuous time Markov process consisting of a sequence of $n$ transient  latent phases and one absorbing state. The Marvov model is shown in Figure~\ref{fig1}~(a).  A CPH distribution is in fact a sequence  of $n$ inter-related Poisson processes, and  the time spent in each latent phase is exponentially distributed.  The process starts in the first phase and progresses sequentially through the other phases with a probability of exiting (to the absorbing state) from any phase.  For  instance,  in a healthcare setting, each latent phase of the Markov model may represent a stage of care and the absorbing state represents a patient exiting hospital. 

Patient LoS may be affected by the patients' information such as gender, age and health condition.  Due to this heterogeneity in the patient population with respect to LoS, a model which makes predictions based on the assumption that all patients have the same LoS distribution  is generally inaccurate \citep{maguire1986}. Therefore, using a CPH model that includes the patient covariate information is necessary as  it identifies the factors which have a significant influence on patient LoS. For example, \cite{faddy1999,faddy2005markov} modelled  the LoS of male geriatric patients in St George's Hospital, London using a Coxian model with two covariates.  They  were included directly into the model parameters (i.e., $\lambda's$ and $\mu's$ displayed in Fig.~\ref{fig1}~(a)). 

Notwithstanding the usefulness of the CPH model in this context, the density function is complicated by the appearance of the matrix exponential, the likelihood surface is multi-modal (cf.~\cite{rizk2019} for further details), and, consequently, parameter estimation can be quite computationally intensive. Including covariates into all model parameters increases the model dimensionality further, and can lead to infeasibly large computational times. \cite{gardiner2012}  modified the standard Coxian model to incorporate covariates on the mean LoS. This method has proven useful in various applications, such as patients with acute myocardial infraction \citep{tang2012}, geriatric patients \citep{marshall2014}, and respiratory patients \citep{zhu2018}.   However, Gardiner only reformulated the  standard Coxian model. The reformulation of  the Coxian model with multiple absorbing states [Fig.~\ref{fig1}~(c)], as well as the joint Coxian and  the conditional Coxian (described briefly in the next paragraph and in more detail in Sections \ref{section2.6} and \ref{section2.7}), will be achieved in this paper. Our reformulation  becomes the basis for our model development for the heterogeneous patient LoS in the ED. 

The emergency department can be seen as a series of service stations that patients which patients pass through before exiting, regardless of the manner by which they exit (discharged home, moved to ward, death, or  left due to impatience).  A station may represent a spell of care (triage, clinician diagnosis, treatment) or simply a waiting room.  Modelling the ED LoS with a CPH distribution may not be suitable as it makes the assumption that each observed station constitutes  one  phase with exponentially distributed waiting time in that particular phase. In reality, each observed station may comprise multiple unrecorded sub-processes (i.e., latent phases), and, in addition, the  LoS distribution is unlikely to be the same in each of the observed stations.  Thus, modelling the ED LoS requires  a  multi-compartment model, where each observed compartment (station) can be modelled using a CPH distribution with a (possibly station-dependent) number of latent phases; this is also known as a joint Coxian model as it specifies a joint distribution for the LoS times in each compartment (see Section 2.5).  This  approach  was used by
\cite{faddy1993} to model the retention time of a drug injected into an organ using a two-compartment model (drug diffusion in and clearance from the body) where each compartment was modelled with a generalised Erlang distribution.  \cite{xie2005jointcox} modelled the LoS of geriatric patients in residential and nursing home care using a two-compartmental model, where the components were time spent in residential and nursing home care, respectively.  \cite{gordon2016} adapted the joint Coxian model by conditioning the LoS in one compartment on the LoS in the previous compartment; this is called the conditional Coxian (see Section \ref{section2.7}). Each compartment was then  modelled with  a conditional CPH distribution. However, a limitation of the models considered by those authors is that covariate influence was not considered. The inclusion of covariates even in the basic Coxian model is already computationally challenging as discussed in \citep{rizk2019}. The compartmental model of course suffers from the same issues (but is even more complex still), and, perhaps, this is the reason that covariates have not been considered previously in the literature.

 In this work, we reformulate the joint Coxian and the conditional Coxian models used in \citep{gordon2016} through a finite mixture of density functions. In this formulation, the inclusion of  covariates becomes straightforward and the numerical calculation of the matrix exponential is avoided, speeding up the fitting process. The data analysed in this paper are taken from the emergency department of University Hospital Limerick (UHL), Ireland from the period December 2016 - August 2017. We analysed the lengths of stay for 37,206 ED patients along with covariates: time of admission, mode of admission, age and, sex. The new methodology is applied to describe the variation in the duration times in the different stations of the ED and assess the effects of the covariates. This study will assist the ED managers in identifying patients who are most likely to have extreme length of stay with respect to the different stations of the ED. 
 
\section{Methodology}
\subsection{Background to Coxian phase-type distribution}\label{section2.1}
 CPH distributions  are a subclass of phase-type (PH) distributions. In recent decades, most researchers have avoided using general PH distributions because they are overparametrised. They are highly redundant as the number of model parameters is greater than the degrees of freedom of the distribution function. The representation of an $n$-PH distribution ($n$ is the number of phases) has in general $n^2+n$ parameters, and its corresponding distribution function has $2n-1$ degrees of freedom \citep{cumani1982}. Using an $n$-CPH distribution reduces the number of parameters to $2n-1$, which makes it non-redundant, while typically still providing an excellent fit to the data. As presented in Figure~\ref{fig1}~(a), the $\lambda$ parameters describe the transition rates through the transient states, while the $\mu$ parameters describe the transition rates from the transient states to the absorbing state.  Furthermore, CPH distributions have the ability to offer superior fit compared to the alternative distributions such as lognormal, Weibull, Gamma, Pareto, or Burr distributions \citep{faddy2009, marshall2014}. 
 
 To estimate the parameters, we use the  maximum likelihood approach. The most used numerical optimisation techniques  to minimise the CPH log-likelihood function, are the expectation-maximization  algorithm \citep{dempster1977} and  the Nelder-Mead simplex method \citep{nelder1965}.  These optimisations methods are numerical and require initiation from a variety of initial values. For more details see \cite{rizk2019} and references therein. 
 
  To obtain the optimal number of latent phases in each station, we fit sequentially an increasing number of phases \citep{faddy1998}, starting with one phase, until little improvement in the fit to the data can be obtained by adding a new phase. The number of phases is determined by minimising the Akaike or the Bayesian information criteria (\textsc{AIC} and \textsc{BIC}).

 \subsection{Coxian phase-type distribution in Matrix form}\label{section2.2}
A CPH distribution is defined as follows: consider a finite and continuous time Markov process $\{ X(t); t \geq  0\}$ with discrete latent transient states $\{1, \dots, n \}$. 
Since the process starts in the first phase, let the row vector $\mathbf{p}=(p_1,\dots,  p_n)=(1,0, \dots, 0)$ be the probability of starting in the transient state $k$, for $k=1,\dots, n$.  Let the column vector $\mathbf{q}=(\mu_1, \dots, \mu_n)^T$ be the absorbing rate  vector, where $\mu_k\geq 0$ is the rate of absorption to the absorbing state from state $k$. The phase-type generator, $\mathbf{Q}$, of the process  is an upper bidiagonal matrix given by 
\begin{align}
\mathbf{Q}=\begin{bmatrix}
        -(\lambda_1 + \mu_1) & \lambda_1 & \quad 0  & \quad \quad&\cdots & \quad 0& \quad 0\\
      0 &  -(\lambda_2 + \mu_2) & \quad\lambda_2 &  &\cdots &\quad 0& \quad 0\\
      \vdots & \vdots & \quad \vdots& & &\quad \vdots&\quad\vdots\\
      0 &  0 & \quad 0 & & \cdots  & \quad -(\lambda_{n-1} + \mu_{n-1})  & \quad \lambda_{n-1}\\
       0 & 0 & \quad 0 &&\cdots  &\quad 0& \quad -\mu_n 
       \end{bmatrix}. \label{matrixQ}
\end{align}

The column vector $\mathbf{q}$ can be written as 
\begin{equation}
\mathbf{q} = -\mathbf{Q}\mathbbm{1},\label{vecq}
\end{equation}
where $\mathbbm{1}$ is an $n$-dimensional column vector of ones.

We denote by $T$ the random variable representing  the time until absorption. The  density function of an CPH distribution with $n$ transient phases is  given by 

\begin{equation}
f(t|\Lambda)=\mathbf{p}\exp(\mathbf{Q}t)\mathbf{q} \label{densityInMatrix}
\end{equation} 

where, $\Lambda=(\lambda_1, \dots, \lambda_{n-1}, \mu_1, \dots, \mu_n)$ is the set of parameters.
The unconditional mean is 
\begin{equation*}
E\big[T\big]=-pQ^{-1}\mathbbm{1} 
\end{equation*}

The matrix exponential, $\exp(\mathbf{Q}t)=\sum_{r=0}^{\infty}(\mathbf{Q}t)^r/r!$, is evaluated numerically. More details on computing matrix exponentials can be found in \citep{moler1978}. The time spent in phase $k$ ($k=1, \dots, n$), denoted by $T_k$, is a random variable that is the minimum of two independent exponential random variables with parameters $\lambda_k$ and $\mu_k$. The random variable $T_k$ is in turn exponential with rate $\theta_k=\lambda_k+\mu_k$, which is also the hazard rate for sojourn in phase $k$ . The expected LoS in  phase $k$  is the reciprocal of the hazard rate in that phase, and is given by

\begin{equation}
LoS_k=1/(\lambda_k +\mu_k)=1/\theta_k, \label{losK}
\end{equation}
where, by definition, $\lambda_n \equiv 0$. 
 
The probability  of exiting phase $k$ to the absorbing state is given by
\begin{align}
\pi_1=&\frac{\mu_1}{\lambda_1+\mu_1}, \> \text{and} \> \> \pi_k=\frac{\mu_k}{\lambda_k + \mu_k}.\prod_{r=1}^{k-1} \Big(\frac{\lambda_r}{\lambda_r+\mu_r} \Big),\quad  k=2,\dots,n. \label{pk}
\end{align}


Despite their wide use in everyday applications, CPH distributions possess the drawback of being computationally intensive to fit to data due to the appearance of the matrix exponential in (\ref{densityInMatrix}); this, in turn, makes the extension to incorporating covariate effects challenging. However, rewriting the model as a mixture of densities  reduces the computational times dramatically and, furthermore, allows the inclusion of covariates, as will be shown in the next subsection. 

\begin{figure}[t]
\centering
\begin{minipage}{.55\textwidth}
  \centering
    \vspace{0.75cm}
  \begin{tikzpicture}
        \tikzset{node style/.style={state, 
                                    fill=white!40!white,
                                    rectangle}}
    
        \node[node style]               (I)   {$1$};
        \node[node style, right=of I]   (II)  {$2$};
       
        \node[draw=none,  right=of II]   (k-n) {$\cdots$};
        \node[node style, right=of k-n]  (n)   {$n$};
        
        \node[draw=none, left=of I] (above) {};
        \node[node style,rectangle,rounded corners=5pt,fill=white!60!white, minimum height=0.4cm,minimum width=0.5cm, below right =1.5cm and 1.3cm] (q1) {Absorbing state};
    \draw[>=latex,auto=left,every loop]
         (above)   edge node {}        (I) 
       (I)  edge [left] node {$\mu_{1}$} (q1)
      (II)  edge node {$\mu_{2}$}  (q1)
       (n)   edge node {$\mu_{n}$}   (q1)
      
  (I) edge node {$\lambda_1$}(II)
  (II)  edge node {$\lambda_2$}(k-n)
   (k-n)  edge node {$\lambda_{n-1}$}  (n);
\end{tikzpicture}
  \caption*{(a)}

\end{minipage}%
\begin{minipage}{.55\textwidth}
  \centering
  \includegraphics[width=1\linewidth]{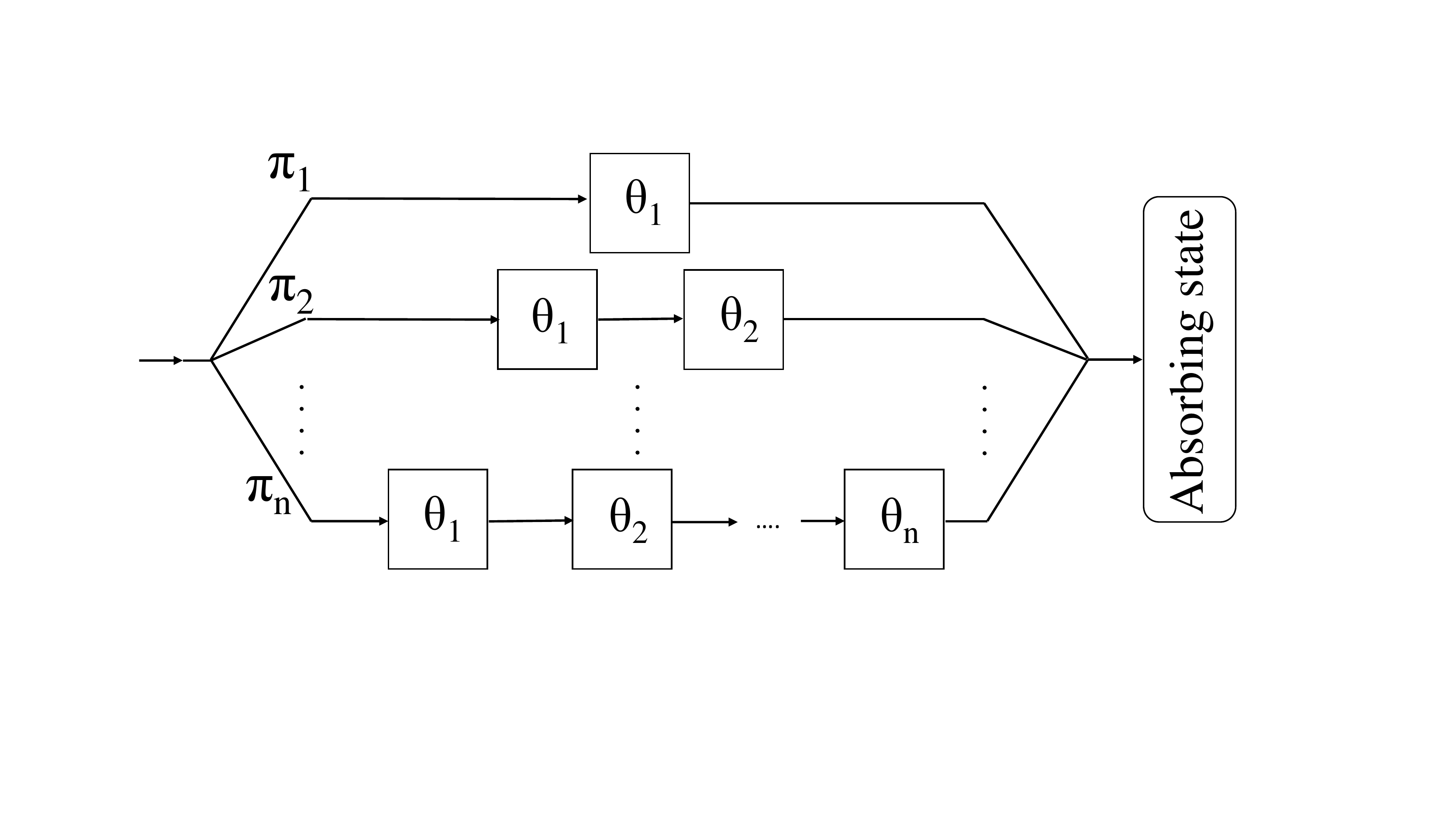}
  \vspace{-1.5cm}
  \caption*{(b)}
\end{minipage}
\end{figure}

  \begin{figure}[t]
\centering
\begin{minipage}{0.55\textwidth}
  \centering
  \begin{tikzpicture}
        \tikzset{node style/.style={state, 
                                    fill=white!40!white,
                                    rectangle}}
    
        \node[node style]               (I)   {$1$};
        \node[node style, right=of I]   (II)  {$2$};
       
        \node[draw=none,  right=of II]   (k-n) {$\cdots$};
        \node[node style, right=of k-n]  (n)   {$n$};
        
        \node[draw=none, left=of I] (above) {};
        \node[node style,rectangle,rounded corners=5pt,fill=white!60!white, minimum height=0.4cm,minimum width=0.5cm, below right =1.65cm and 1.3cm] (q1) {Absorbing state 1};
         \node[node style,rectangle,rounded corners=5pt,fill=white!60!white, minimum height=0.4cm,minimum width=0.5cm, above right =1.65cm and 1.3cm] (q2) {Absorbing state 2};
    \draw[>=latex,auto=left,every loop]
         (above)   edge node {}        (I) 
       (I)  edge [left] node {$\mu_{11}$} (q1)
      (II)  edge node {$\mu_{21}$}  (q1)
       (n)   edge node {$\mu_{n1}$}   (q1)
       (I)  edge node {$\mu_{12}$} (q2)
      (II)  edge [right] node {$\mu_{22}$}  (q2)
       (n)   edge [right] node {$\mu_{n2}$}   (q2)
  (I) edge node {$\lambda_1$}(II)
  (II)  edge node {$\lambda_2$}(k-n)
   (k-n)  edge node {$\lambda_{n-1}$}  (n);
\end{tikzpicture}   
  \caption*{(c)}

\end{minipage}%
\begin{minipage}{.55\textwidth}
  \centering
  \includegraphics[width=1\linewidth]{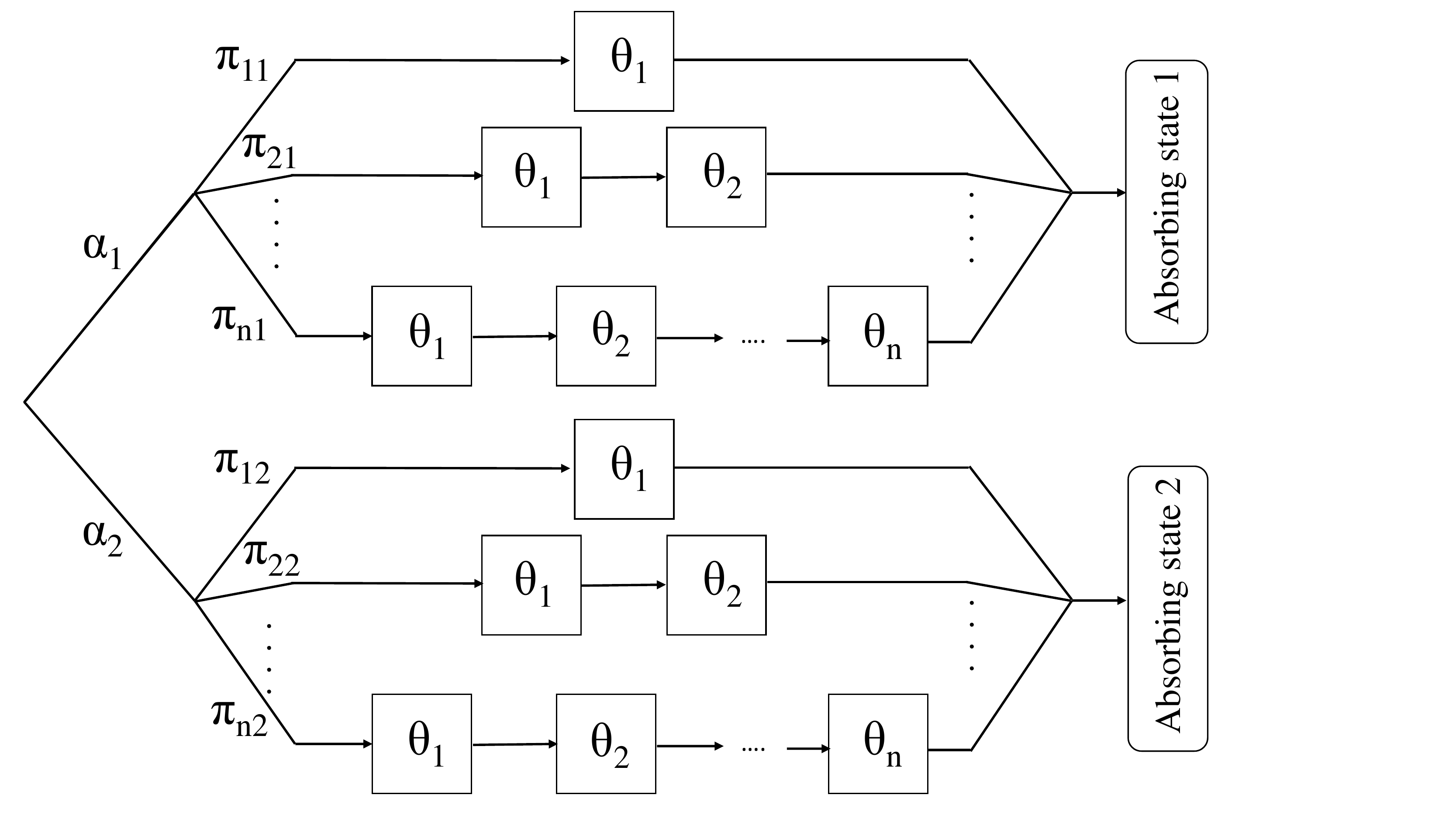}
  \caption*{(d)}
\end{minipage}
\caption{(a) An illustration of the Coxian Markov Model with $n$ phases. (b) Equivalent arrangement of an n-phase CPH distribution.  (c) Representation of a Coxian model with two absorbing states. (d) Equivalent arrangement of a Coxian model with two absorbing states. \label{fig1}}
\end{figure}

\subsection{Coxian phase-type distribution in mixture form}\label{section2.3}
Based on the generalisation of Erlang's method of stages \citep{erlang1917}, a CPH distribution can  be expressed as a mixture of densities. First of all, we note that a parallel connection of $n$ exponential random variables is modelled as a mixture of the $n$ exponential distributions. Each of the  mixture components, $\pi_k$, $k=1, \dots, n$, represents the probability of choosing route $k$. In a similar fashion, the Coxian Markov model of $n$ transient phases can be aggregated into $n$ parallel routes as shown in Fig.~\ref{fig1}~(b). Each route  is a series or  convolution of $k$ independent exponential random variables with rates $\theta_1, \dots, \theta_k$, yielding a hypoexponential distribution whose mean is, of course, $1/\theta_1 + \cdots + 1/\theta_k$. Therefore, a CPH  distribution with $n$ phases can be written as a mixture of $n$ hypoexponential distributions. Then, avoiding the matrix form in (\ref{densityInMatrix}), the density can be written as

\begin{equation}
f(t|\Theta)=\sum_{k=1}^n\pi_k h_k(t), \label{density2}
\end{equation}
where $\Theta=(\pi_1, \dots, \pi_{n-1}, \theta_1, \dots, \theta_n)$ is the set of parameters, $\sum_{k=1}^n\pi_k=1$, and  $h_k(t)$ is the density of a hypoexponential distribution given by
\begin{equation}
h_k(t)=\sum_{r=1}^k \Bigg(\prod_{\underset{c\neq r}{c=1}}^k \frac{\theta_c}{\theta_c-\theta_r}\Bigg) \theta_r e^{-\theta_rt}.  \label{h_k}
\end{equation}

The unconditional mean absorption time of a CPH distribution in terms of the new parameters, $\theta$ and $\pi$, is
\begin{equation*}
E[T]=\pi_1\frac{1}{\theta_1}+ \pi_2(\frac{1}{\theta_1}+\frac{1}{\theta_2}) + \dots + \pi_n(\frac{1}{\theta_1}+ \dots + \frac{1}{\theta_n}).
\end{equation*}

If all the $\theta$ parameters are equal, the expression defined in (\ref{density2}) reduces to an Erlang density, and the CPH density function becomes a mixture of Erlang densities. If two or more $\theta$'s are equal, alternative expressions for the CPH distribution can be obtained by inverting its Laplace transform,  which is given by 
\begin{equation}
f^*(s)=\mathscr{L}\{f(t)\}= \sum_{k=1}^n\pi_k Z_k(s),\label{laplace}
\end{equation}
where 
\begin{equation*}
Z_k(s)=\mathscr{L}\{h_k(t)\}=\frac{\theta_1\theta_2 \dots \theta_k}{(\theta_1 + s)(\theta_2+s) \dots (\theta_k + s)}
\end{equation*}
is the Laplace transform of a hypoexponential distribution with $k$ phases. Equation~(\ref{laplace}) is a weighted sum of rational expressions in $s$ with numerators of degree zero.  This structure allows straightforward inversion  based on partial fraction decomposition. However in practice, when estimating the parameters by numerically optimising the loglikelihood function, it is highly unlikely to encounter numerically identical $\theta$'s; thus, Equation~(\ref{density2}) is often sufficient as the density function for practical purposes (without the need for Laplace transformation).\\

The rates, $\theta_1, \dots, \theta_n$ and the mixture components, $\pi_1, \dots, \pi_n$, are the same as defined in Section~\ref{section2.2}, where the mean length of stay in phase $k$, $k=1, \dots, n$, is LoS$_k=1/\theta_k$,  and the absorption  probability from phase $k$ is $\pi_k$.  Instead of the matrix representation of the CPH model that involves   transient and absorption parameters, $(\lambda_1, \dots, \lambda_n, \mu_1, \dots, \mu_n)$, we now have a new representation with  parameters $(\pi_1, \dots, \pi_n, \theta_1, \dots, \theta_n)$. The number of parameters $2n-1$ remains the same   since $\pi_n=1-\sum_{k=1}^{n-1}\pi_k$.  Furthermore, if required, the transient rates, $\lambda$, and the absorbing rates, $\mu$, can be retrieved from the following recurrence formula, which follows from Equations~(\ref{losK})~and~(\ref{pk}),

\begin{align*}
\lambda_1=&\theta_1 - \mu_1 \quad \& \quad \mu_1= \pi_1 \theta_1, \quad \text{for} \quad  k=1,\\
\lambda_k=&\theta_k - \mu_k \quad \& \quad \mu_k= \pi_k\frac{\prod_{r=1}^k \theta_r}{\prod_{r=1}^{k-1}\lambda_r},\quad  \text{for}\> \> k=2,\dots, n.
\end{align*}
 
\subsection{Coxian distribution with multiple absorbing states}\label{section2.4}
The ED system consists of a series of service stations.  After going through a station, which consists of latent phases of care, the patient will either exit the system or proceed to other stations. Each station can be modelled with a CPH distribution. However, to model the movement between stations, it is necessary to include an additional absorbing state: one absorbing state represents  patient leaving the ED, and  the other absorbing state represents  movement to the next ED station.  Figure~\ref{fig1}~(c) depicts a CPH model with two absorbing states.  An additional absorbing state can  be incorporated by modifying the matrix $\mathbf{Q}$ and the vector $\mathbf{q}$ from the basic CPH distribution \citep{garg2009clustering, mcclean2010using}. Suppose we have a Coxian model with $n$ transient phases and two absorbing states. The vector $\mathbf{q}$ no longer satisfies Eq.~(\ref{vecq}) and it becomes an $n\times 2$ matrix defined as 
\begin{align*}
\mathbf{q} =\big[\mathbf{q_1}, \mathbf{q_2} \big]=\begin{bmatrix}
       \mu_{11} & \mu_{12} \\
       \mu_{21} & \mu_{22} \\
      \vdots &  \vdots\\
      \mu_{n1} & \mu_{n2}
       \end{bmatrix},
\end{align*}
where $\mathbf{q_1}$  and $\mathbf{q_2}$ are the transition rate vectors  to the first  and second absorbing states respectively. The matrix $\mathbf{Q}$ becomes 
\begin{align*}
\mathbf{Q}=\begin{bmatrix}
        -(\lambda_1 + \mu_{11}+ \mu_{12} ) & \lambda_1  & \quad &\cdots &  \quad 0\\
      0 &  -(\lambda_2 + \mu_{21}+ \mu_{22}) &  &\cdots & \quad 0\\
      \vdots & \vdots & &  &\quad \vdots\\
       0 & 0  &&\cdots  & \quad -(\mu_{n1}+\mu_{n2})
       \end{bmatrix}. \label{matrixQ}
\end{align*}
In matrix form, the density function of the CPH distribution with two absorbing states is a vector given by 
\begin{equation}
\mathbf{f}(t|\Lambda)=\{f_1(t),f_2(t)\}=\big\{\mathbf{p}\exp(\mathbf{Q}t)\mathbf{q_1}, \mathbf{p}\exp(\mathbf{Q}t)\mathbf{q_2}\big\},\label{multipleAbsStateMatrix}
\end{equation}
where $\mathbf{p}=(1,0, \dots, 0)$ and $\Lambda=(\lambda_1, \dots, \lambda_{n-1},\mu_{11}, \dots, \mu_{n1}, \mu_{12}, \dots, \mu_{n2})$.

In previous research, only the matrix form of a CPH distribution with multiple absorbing states was presented. We derive here the mixture form. 

Since the system has two exits, the Markov  model can be rearranged into two different routes, where each route leads to a Coxian Markov chain with one absorbing state as shown in Figure~\ref{fig1}~(d).
The probability density function of the LoS prior to absorption into the various absorbing states is a vector of two Coxian densities: (i) the density of the LoS prior to exiting the system (the ED in our case), we denote by $f_1(t)$, and (ii) the density of the LoS prior to continuing  to the next station of the ED and we denote by $f_2(t)$. The density function of the CPH distribution with two absorbing states is then given by 
\begin{equation}
\mathbf{f}(t|\Theta)=\{f_1(t),f_2(t)\}=\Big\{\sum_{k=1}^n \pi_{k1} h_k(t), \sum_{k=1}^n \pi_{k2} h_k(t)\Big\},\label{multipleAbsState}
\end{equation}
where $\Theta=(\pi_{11}, \dots, \pi_{n1}, \pi_{12}, \dots, \pi_{n2} , \theta_1, \dots, \theta_n)$ is the set of parameters, $\sum_{k=1}^n \pi_{k1}+ \sum_{k=1}^n \pi_{k2}=1$, and $ h_k(t)$ is as defined in Equation~(\ref{h_k}).

The unconditional expected LoS prior to absorption in each absorbing state is given by the vector $\mathbf{E}=(E[T_1], E[T_2])$, where 
\begin{equation*}
E[T_j]=\pi_{1j}\frac{1}{\theta_1}+ \pi_{2j}(\frac{1}{\theta_1}+\frac{1}{\theta_2}) + \dots + \pi_{nj}(\frac{1}{\theta_1}+ \dots + \frac{1}{\theta_n}), \quad j=1,2.
\end{equation*}
 We define $\alpha_i$, to be the probability that the $i$th observation in the data has taken one of the two routes. However, in practice, we observe the station from which an individual exits, and,  therefore, $\alpha_i$  simply becomes an indicator variable, describing which of the two events has occurred for each observation (exit the system or exit to the next station), i.e.,

\[   \alpha_i= \left\{
\begin{array}{ll}
      1 & \text{the individual} \> \> i \> \> \text{has exited to the first  absorbing state} \\
      0 & \text{otherwise} \\
\end{array}
\right.. \] \label{alpha}

Then, the log-likelihood of an observed route visited and duration spent for an individual can be written as
\begin{equation*}
\ell(\Theta,\alpha,\mathbf{t})= \sum_i \log\Big\{\alpha_i f_1(t_i|\Theta) + (1-\alpha_i) f_2(t_i|\Theta)\Big\}.
\end{equation*}

The transient rates, $\lambda_1, \dots, \lambda_n$, and the absorbing rates, $\mu_{11}, \dots, \mu_{n1}, \mu_{12}, \dots, \mu_{n2}$, can be obtained recursively  as follows,
\begin{align*}
\lambda_1&=\theta_1 - \mu_{11}-\mu_{12},  \quad \mu_{11}=\pi_{11}\theta_1 \quad  \& \quad \mu_{12}=\pi_{12}\theta_1, \quad \text{for} \quad k=1, \\
\lambda_k&=\theta_k - \mu_{k1}-\mu_{k2}, \quad \mu_{k1}= \pi_{k1}\frac{\prod_{r=1}^k \theta_r}{\prod_{r=1}^{k-1}\lambda_r} \quad \& \quad  \mu_{k2}= \pi_{k2}\frac{\prod_{r=1}^k \theta_r}{\prod_{r=1}^{k-1}\lambda_r}, \quad  \text{for}\quad k=2,\dots, n.
\end{align*}
\subsection{Computational speed}\label{section2.5}
 Unlike the matrix form, the mixture representations of the density functions do not contain the matrix exponential term. This makes the data fitting process  much faster and allows fitting large data.  
Note that the number of parameters increases from $2n-1$ in the standard Coxian, to $3n-1$ in the Coxian with two absorbing states.  Table~\ref{table1}  shows a comparison between the fitting times of the matrix and the mixture representations. We simulated two 3-phase CPH distributions, one with one absorbing state  and another with two absorbing states and used two different sample sizes. We fitted the distributions in MATLAB \citep{MATLAB2015} on a PC with a 3.00 GHz processor. As shown in the table, the computational times drop dramatically with the mixture form.

\begin{table}[t]
\caption{Fitting times in seconds of a 3-phase CPH distribution in matrix and mixture forms \label{table1}}
 \centering
 \resizebox{\columnwidth}{!}{
\begin{tabular}{cccc|ccc} 
 \hline
&\multicolumn{3}{c}{\bf One absorbing state}& \multicolumn{3}{c}{\bf Two absorbing states} \\
 \hline
 \bf Sample size & \bf Matrix form &  \bf Explicit form & \bf Relative speed &\bf Matrix form &  \bf Explicit form &\bf Relative speed  \\ 
\hline
1,000&    132 s &0.6 s & 220 &  204 s &  1.6 s & 127.5\\
\hline
5,000 &  516 s &   0.8 s & 645 & 1200 s	&  3.5 s & 342.9\\
\hline
\end{tabular}
 }
\end{table}

\subsection{Joint Coxian distributions}\label{section2.6}
As mentioned previously, the ED consists of a series of stations where each can be modelled with a Coxian distribution with two absorbing sates. In a system of total of $N$ stations, a patient who spends a time $t_L$ in a particular station $S_L$ ($L=1, \dots, N$), is the same patient who already had lengths of stay $t_1, \dots, t_{L-1}$ in the preceding stations $S_1, \dots, S_{L-1}$, respectively. 

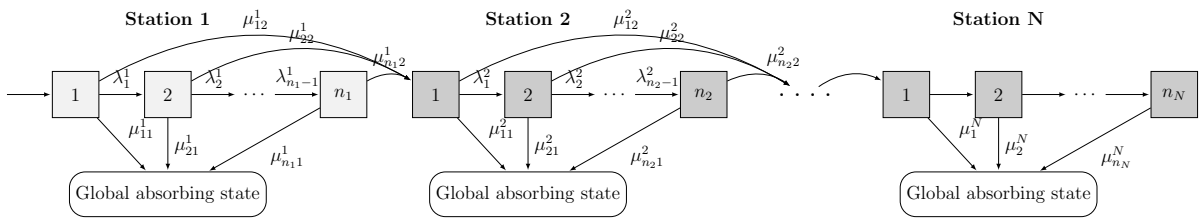
\begin{figure}[b]
\centering
 
 \resizebox {\columnwidth} {!} {
\begin{tikzpicture}
        \tikzset{node style/.style={state, 
                                    fill=white!95!black,
                                    rectangle}}
    
        \node[node style]               (Ia)   {$1$};
        \node[node style, right=of Ia]   (IIa)  {$2$};
        \node[draw=none,  right=of IIa]   (k-na) {$\cdots$};
        \node[node style, right=of k-na]  (na)   {$n_1$};
        
        \tikzset{node style/.style={state, 
                                    fill=white!80!black,
                                    rectangle}}
          \node[node style, right=of na]  (Ib)   {$1$};
        \node[node style, right=of Ib]   (IIb)  {$2$};
        \node[draw=none,  right=of IIb]   (k-nb) {$\cdots$};
        \node[node style, right=of k-nb]  (nb)   {${n_2}$};
        
        \tikzset{node style/.style={state, 
                                    fill=white!60!black,
                                    rectangle}}
       \node[draw=none, right=of nb]   (Ic)  {{\bf . . . .} };
        
        \tikzset{node style/.style={state, 
                                    fill=white!80!black,
                                    rectangle}}
          \node[node style, right=of Ic]  (Ie)   {$1$};
        \node[node style, right=of Ie]   (IIe)  {$2$};
        \node[draw=none,  right=of IIe]   (k-ne) {$\cdots$};
        \node[node style, right=of k-ne]  (ne)   {${n_N}$};

       \node[draw=none, left=of Ia] (left) {};
       \node[draw=none,  above=of IIa,outer sep=-0.1cm]   (station1) {\bf{Station 1}};
       \node[draw=none,  above=of IIb,outer sep=-0.1cm]   (station2) {\bf{Station 2}};
        \node[draw=none,  above=of IIe,outer sep=-0.1cm]   (station4) {\bf{Station $\bf N$}};
       
        \node[node style,rectangle,rounded corners=10pt,fill=white!60!white,minimum height=1cm,minimum width=3.5cm,node distance=1.2cm, below=of IIa] (qa) {Global absorbing state};
        \node[node style,rectangle,rounded corners=10pt,fill=white!60!white,minimum height=1cm,minimum width=3.5cm,node distance=1.2cm, below=of IIb] (qb) {Global absorbing state};
        \node[node style,rectangle,rounded corners=10pt,fill=white!60!white,minimum height=1cm,minimum width=3.5cm,node distance=1.2cm, below=of IIe] (qe) {Global absorbing state};
        
       \draw[>=latex,auto=left,every loop]
         (left)   edge node {}        (Ia) 
           (Ia) edge node {$\lambda^1_1$}(IIa)
  (IIa)  edge node {$\lambda^1_2$}(k-na)
   (k-na)  edge node {$\lambda^1_{n_{1}-1}$}  (na)
       (Ia)  edge node {$\mu^1_{11}$} (qa)
      (IIa)  edge node {$\mu^1_{21}$}  (qa)
       (na)   edge node {$\mu^1_{n_{1}1}$}   (qa)
        (Ia)  edge [bend left]  node {$\mu^1_{12}$} (Ib)
      (IIa)  edge [bend left]  node {$\mu^1_{22}$}  (Ib)
       (na)   edge [bend left]  node {$\mu^1_{n_12}$}   (Ib)
     
   (Ib) edge node {$\lambda^2_1$}(IIb)
  (IIb)  edge node {$\lambda^2_2$}(k-nb)
   (k-nb)  edge node {$\lambda^2_{n_{2}-1}$}  (nb)
       (Ib)  edge node {$\mu^2_{11}$} (qb)
      (IIb)  edge node {$\mu^2_{21}$}  (qb)
       (nb)   edge node {$\mu^2_{n_{2}1}$}   (qb)
       (Ib)  edge [bend left]  node {$\mu^2_{12}$} (Ic)
      (IIb)  edge [bend left]  node {$\mu^2_{22}$}  (Ic)
       (nb)   edge [bend left]  node {$\mu^2_{n_{2}2}$}   (Ic)

(Ic)  edge [bend left]  node {} (Ie)
             (Ie) edge node {}(IIe)
  (IIe)  edge node {}(k-ne)
   (k-ne)  edge node {}  (ne)
       (Ie)  edge node {$\mu^{N}_{1}$} (qe)
      (IIe)  edge node {$\mu^{N}_{2}$}  (qe)
       (ne)   edge node {$\mu^{N}_{n_{N}}$}   (qe);

\end{tikzpicture}
}   
\caption{Example of a system consisting of successive service stations. \label{Fig_seriesCox}}
\end{figure}

To take into account the movement between and within the ordered sequence of stations, \cite{xie2005jointcox} used a joint probability density function to model the LoS in a successive types of care which is similar to the scenario outlined in this research. The density is derived from the work of \cite{fredkin1986} on aggregated Markov processes.  Note that, each station, apart from the last one, has two absorbing states: the global absorbing state (exiting the ED), and the first phase of the next station as shown in Figure~\ref{Fig_seriesCox}. We define two absorbing rates vectors for each station: $\mathbf{q_{L1}}$ represents exiting from station $S_L$ to the global absorbing state, and $\mathbf{q_{L2}}$ represents exiting from station $S_L$ to the next station $S_{L+1}$.  
In matrix form, the density function of the joint system of $N$ stations is 
\begin{align*}
\mathbf{g}(t)=\sum^N_{L=1}\gamma^L_i g_L(t),
\end{align*}
where,
\begin{align}
g_L(t)=&f(t_1 \cap t_2 \cap \dots \cap t_L) \nonumber \\
=&\mathbf{p}_{1}\exp{(\mathbf{Q}_{1}t_1)}\mathbf{T}_{12}\exp{(\mathbf{Q}_{2}t_2)}\mathbf{T}_{23}\exp{(\mathbf{Q}_{3}t_3)} \dots \mathbf{T}_{L-1,L}\exp{(\mathbf{Q}_{L}t_L)}\mathbf{q}_{L1}, \label{jointdensity}
\end{align}
is the density  for a patient undergoing absorption to the global absorbing state from station $L$, and $\gamma_i^L$ is a dummy variable indicating which station the individual exited from, e.g., for $L=1$, $g_1(t)=\mathbf{p}_{1}\exp{(\mathbf{Q}_{1}t_1)}\mathbf{q}_{11}$.

The matrix $T_{m,m+1}$ is defined as 
\begin{align*}
T_{m,m+1}=\begin{bmatrix}
        \mu^{m}_{12} & 0 & \cdots & 0\\
      \mu^{m}_{22} & 0 & \cdots & 0\\
      \vdots & \vdots & \ddots&  \vdots \\
       \mu^{m}_{n_{m}2} & 0 & \cdots & 0
       \end{bmatrix},
\end{align*}
for , $m=1, \dots, L-1$. It is of dimension $n_{m}\times n_{m+1}$ where $n_{m}$ and $n_{m+1}$ are the number of phases in stations $S_{m}$ and $S_{m+1}$ respectively.  This matrix represents patients transferring from a particular station to a successive one. The reason  $\mathbf{T}_{m,m+1}$ contains non-zero elements in the first column, with all other elements equal to zero, is due the fact that patients may only enter the first phase of station $S_{m+1}$  from any of the transient phases of the preceding station $S_{m}$. The vector $(\mu^{m}_{12}, \dots, \mu^{m}_{n_{m}2})^T$, the first column of matrix $\mathbf{T}_{m,m+1}$, is  the absorbing rate vector, $\mathbf{q}_{m2}$, defined above. It represents the individuals exiting from station  $S_{m}$ to the first phase of station $S_{m+1}$; this is "absorbing" from the perspective of station $S_m$ as the patient cannot return to this station, but, of course, the patient has not exited the system to the global absorbing state.

Expression~(\ref{jointdensity}) can be simplified further. We can in fact write it as a product of $L$ Coxian densities. We achieve this due to the fact that  the matrix  $\mathbf{T}_{m,m+1}$ is the outer product of the $n_m\times1$ absorbing rate vector, $\mathbf{q}_{m2}$, from station $S_{m}$ to $S_{m+1}$, and the $1\times n_{m+1}$ initial probability vector, $\mathbf{p}_{m+1}$, for station $S_{m+1}$,
\begin{align*}
\mathbf{T}_{m,m+1}=(\mu^{m}_{12}, \mu^{m}_{22}, \cdots, \mu^{m}_{n_{m}2})^T\otimes (1, 0,  \cdots, 0)=\mathbf{q}_{m2}\otimes \mathbf{p}_{m+1}.
       \end{align*}
The joint probability in (\ref{jointdensity}) then becomes 
\begin{align}
 g_L(t)=\mathbf{p}_{1}\exp{(\mathbf{Q}_{1}t_1)}\mathbf{q}_{12}\otimes \mathbf{p}_{2}\exp{(\mathbf{Q}_{2}t_2)}\mathbf{q}_{22}\otimes \dots \otimes \mathbf{p}_{L}\exp{(\mathbf{Q}_{L}t_L)}\mathbf{q}_{L1}.  \label{jointdensity2}
 \end{align}
 This is Bayes' theorem for $L$ events, 
 \begin{align*}
g_L(t)=f(t_1 \cap t_2 \cap \dots \cap t_L)=f\{t_1|(t_2 \cap \dots \cap t_L)\}\times f\{t_2|(t_3 \cap \dots \cap t_L)\}\times \dots \times f(t_L),
 \end{align*}
 
 where $f\{t_{m}|(t_{m+1} \cap \dots \cap t_L)\}=\mathbf{p}_{m}\exp{(\mathbf{Q}_{m}t_{m})}\mathbf{q}_{m2}$, is the density for the time spent in station $S_m$, given that the individual proceeded through each subsequent station exiting from station $S_L$.

Writing the density in this new form, (\ref{jointdensity2}), allows us to replace each component of the product by its equivalent mixture form defined in Sections \ref{section2.3} and \ref{section2.4}, leading to a density function without any matrix exponential terms. Nevertheless, this approach requires simultaneously calculating  the joint probability over each station, resulting in a large number of parameters to be estimated. This may lead to instability in the likelihood optimisation process, especially in the case of large number of stations.  Furthermore, when including covariates, estimating the parameters may become computationally infeasible. Despite using the matrix form, this methodology was successful in \citep{xie2005jointcox}  due the low number of stations (two) used in the model as well as the low number of latent phases in each station (one phase and two phases respectively). In addition, the sample size was reasonably small (935 observations) and the model did not include covariates. In the following section, we present an alternative approach where  the system can be considered using two consecutive stations at a time.

\subsection{Conditional Coxian  distribution}\label{section2.7}
The conditional Coxian phase-type model   was used by \cite{gordon2016} to model patient transitions between hospital and community. The model is fitted over  two consecutive stations at a time. The probability density function for the LoS in a particular station  is conditioned on the LoS observed in the previous station.  Here, the estimated parameters from the first station feed into the estimated parameters for the second station; then, the estimates from the second station are used for the third, and so on.  However, the model in \citep{gordon2016} is given in the matrix form which is computationally expensive. This is perhaps the reason that the authors did not incorporate covariates into the model. In this section, we present the conditional Coxian phase-type distribution in the mixture form.

 Let $S_m$ and $S_{m+1}$ two consecutive stations. The latter is not necessarily the last station. Their corresponding number of transient phases are $n_m$ and $n_{m+1}$ respectively. Each exhibits two absorbing states: (i) the global absorbing state, and (ii) the first phase of the proceeding station (where the first phase of $S_{m+1}$ is the absorbing state of $S_m$). Let  $t_{m+1}$  the length of time spent of an individual at the current station $S_{m+1}$  and $t_{m}$ is the length of time spent of the same individual at the previous station $S_{m}$. Thus, from Bayes' theorem, the conditional CPH model is

\begin{align}
\mathbf{f}(t_{m+1}|t_{m})&= \frac{\mathbf{f}(t_{m} \cap t_{m+1})}{f(t_{m})}= \frac{f(t_{m}|t_{m+1})\times \mathbf{f}(t_{m+1})}{f(t_{m})},
 \label{conddensity1}
\end{align}

 where $f(t_{m}|t_{m+1})=f(t_m |\text{ exit to next station} \> S_{m+1})=f_2(t_{m})$  is the second component of the probability density function for the Coxian distribution with two absorbing states defined in (\ref{multipleAbsState}). It models the patients that spent time $t_{m}$ in $S_m$ given they proceeded to station $S_{m+1}$. The patients who have already left station $S_m$  through the global absorbing state (representing discharge, death, transfer, etc.) are not considered for station $S_{m+1}$.  The vector $\mathbf{f}(t_{m+1})=\{f_1(t_{m+1}), f_2(t_{m+1})\}$ (Eq.~\ref{multipleAbsState}) represents  the probability density function for the Coxian phase-type distribution with two absorbing states at station $S_{m+1}$. Note that, only the second component $f_2(t_{m+1})$ will be considered for the density at the next station right after $S_{m+1}$. The denominator, $f(t_m)$, is the marginal probability density representing the patients who spent time $t_m$ in station $S_m$ before absorption to either of the two absorbing states.

 As a result, and by using the explicit forms defined in Section~\ref{section2.4}, the conditional density in (\ref{conddensity1})  becomes 
\begin{align}
\mathbf{f}(t_{m+1}|t_{m})&=\frac{f_2(t_{m})\times \{f_1(t_{m+1}), f_2(t_{m+1})\}}{f(t_{m})}\\
&=\frac{\sum_{k=}^{n_A} \pi^m_{k2} h^m_k(t_{m})\times \Big\{ \sum_{k=1}^{n_B} \pi^{m+1}_{k1} h^{m+1}_k(t_{m+1}), \sum_{k=1}^{n_{m+1}} \pi^{m+1}_{k2} h^{m+1}_k(t_{m+1})\Big\}}{\sum_{k=1}^{n_A} \pi^m_{k} h_k^{m}(t_{m})}. \label{conddensity}         
\end{align}

 If the station $S_{m+1}$ the last station, then it will only have one absorbing state (global absorbing state).  In this case, the two component vector $\mathbf{f}(t_{m+1})=\{f_1(t_{m+1}), f_2(t_{m+1})\}$ is replaced with one component, $f(t_{m+1})$, representing one absorbing state.

  Suppose we observe event times of $I$ individuals $\mathbf{t_{m+1}}=(t_{1(m+1)}, \dots, t_{I(m+1)})$  and $\mathbf{t_{m}}=(t_{1m}, \dots, t_{Im})$ at stations  $S_{m+1}$ and $S_m$ respectively, then the log-likelihood function of the model at station $S_{m+1}$ will be given by
 \begin{align}
\ell(\Theta^{m+1},\mathbf{t_{m+1}}|\mathbf{t_{m}})&= \sum^I_{i=1}\log\Big\{\frac{f_2(t_{i m}, \Theta^m_2)\times \big[\alpha_i f_1(t_{i (m+1)},\Theta_1^{m+1})+(1-\alpha_i)f_2(t_{i (m+1)},\Theta_2^{m+1})\big ]}{f(t_{i m},\Theta^m)}\Big\}, \label{Logconddensity}      
\end{align}
where, $\Theta^{m+1}=\Theta^{m+1}_1\cup\Theta^{m+1}_2=(\pi^{m+1}_{11}, \dots \pi^{m+1}_{n_{(m+1)}1}, \pi^{m+1}_{12}, \dots, \pi^{m+1}_{n_{(m+1)}2}, \theta^{m+1}_1, \dots, \theta^{m+1}_{n_{m+1}})$  is the set of $3n_{m+1}-1$ parameters to be estimated, and $\alpha_i$ is an indicator variable as defined in Section~\ref{section2.4} . The  parameters  $\Theta^m_2=(\pi^m_{12}, \dots \pi^m_{n_{m}2}, \theta^m_1, \dots, \theta^m_{n_{m}})$ and $\Theta^m=(\pi^m_{1}, \dots, \pi^m_{n_{m}},\theta^m_1, \dots, \theta^m_{n_{m}})$ are  the optimal parameter estimates  from the implementation of the  methodology for the previous station $S_m$.

\subsection{Conditional Coxian with covariates}\label{section2.8}
 The new mixture form of the conditional Coxian density function allows  straightforward inclusion of covariates directly into the distributional parameters, particularly into  $\theta$ the hazard rate parameters. Our primary goal is to assign patients to different LoS groups in each ED station and explain the difference in the expected LoS by covariates.  For the  event times of $I$ individuals, $\mathbf{t_{m+1}}=(t_{1(m+1)}, \dots, t_{N(m+1)})$, observed at station $S_{m+1}$, let $\bf{X=(x_1, \dots, x_N)^T}$ be the covariate information matrix where $\mathbf{x_i}=(x_{1i}, \dots, x_{li})$. The covariates can be incorporated into the distribution  by allowing the hazard rates, $\theta^{m+1}_k$, $k=1, \dots, n_{m+1}$, to depend on them through log-linear functions: $\theta^B_k=\theta^{m+1}_{0k}\exp(-\bf{x_i} \beta^{m+1})$, where $\theta^{m+1}_{0k}$ is the phase-specific intercept $k=1,\ldots,n_{m+1}$, and $\mathbf{\beta^{m+1}}=(\beta^{m+1}_{1},\dots,  \beta^{m+1}_{l})$ is the slope coefficient vector.  The conditional mean time is, $E[T^{m+1}|\mathbf{x_i}]=\exp[b^{m+1}_0+ \exp(\mathbf{x_i} \beta^{m+1})]$, where $b^{m+1}_0=\sum^n_{k=2}\pi^{m+1}_k\big(\sum^k_{r=1}1/\theta^{m+1}_{0r} \big)$. 
  
At the cost of a substantial increase in the number of parameters, the covariates can be incorporated by allowing the slopes to depend on the   phase $k$: $\theta^{m+1}_k=\theta^{m+1}_{0k}\exp(-\bf{x_j}{\bf\beta^{m+1}_k})$. Furthermore, covariates can be even added into the $\pi$ absorption probabilities. However, we do not consider that here, and find that placing covariates only in the $\theta$ parameters is sufficient to provide a very good fit in our application.

\section{Application}
\subsection{The data}
Emergency department admission data for University Hospital Limerick (UHL) were provided by the Health Service Executive (HSE), Ireland.   It includes the following patient information: arrival mode (ambulance/other), arrival date and time, age, sex, triage start time, clinician examination start time, departure date and time, and destination upon departure. The format of patient information collection is shown in Table~\ref{data_format}. Triage is performed by ED nurses, and is a method of sorting patients according to their need for emergency medical attention. 

\begin{figure}[t]
\centering
\includegraphics[width=1\linewidth]{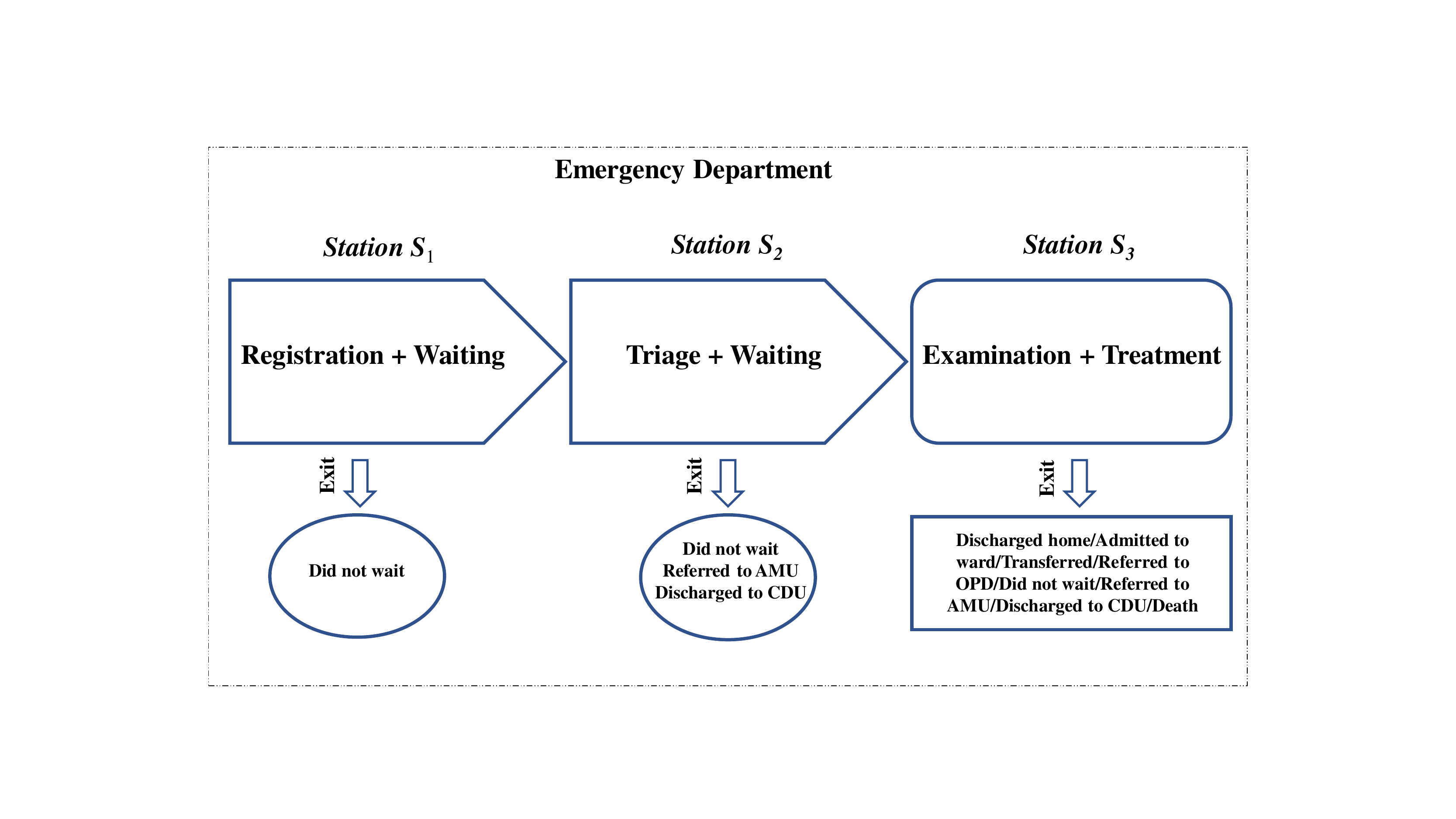}
\vspace*{-2cm}
\caption{Patient flow conceptual model of the ED at UHL. \label{EDModel}}
\end{figure}

 Based on the data provided,  the total waiting time in the ED can be divided into  three successive stations. The first station, $S_1$, is a waiting room, where patients proceed  after registering at the reception desk. They wait to be called for triage.   The length of stay in the second station, $S_2$, is the time spent in triage  plus the waiting time to be seen by the ED clinician. The third station, $S_3$,  is where the clinician examination and the treatment take place. The patient flow model is described in Figure~\ref{EDModel}. Patients may exit the ED from any of the three stations. Those who exit from $S_1$ leave the ED before even entering the triage station, i.e., they are not prepared to wait. Upon completion of triage in station $S_2$, the nurse either (a) transfers the patient to the acute medical unit (AMU) or to the critical decision unit (CDU), or (b) to the waiting room to wait for the ED clinician. The patients who are in the waiting room may also decide to exit the ED without waiting further. Finally, patients proceed to the final station, $S_3$, where they are examined and receive treatment before they exit the ED to various destinations such as: home, ward, outpatient department (OPD), AMU, CDU, death, or simply leave before the start of the treatment. The recorded destination plays an important role in identifying the station from which the patient exited.  For example, the second patient listed in Table~\ref{data_format}, is missing the examination time, and the destination is ``Did not wait'', i.e., this patient exited from station $S_2$ without waiting for treatment.
 
Here we analyse a sample of 37,206 full records from December 2016 to August 2017. Of these patients, 129 exited $S_1$, 2,785 exited $S_2$, and 34,292 exited from $S_3$, i.e., the vast majority flow through the whole system. Furthermore, the covariates are: arrival time,  admission mode (ambulance or other), age, and sex. For the purpose of this study, patient arrival time and age have been discretised. The arrival time is divided into day (08:00:00-19:59:59) and night (20:00:00-07:59:59). Patient age is divided into  $<18$, $18-44$, $45-64$ and $\geq 65$ years respectively. The patients who arrive at night represent $30\%$ of all patients in the dataset. Twenty-six percent of patients arrived by ambulance and approximately half of the patients are females. Patients in the age categories $<18$, $18-44$, $45-64$, and $\geq 65$ represent respectively $26\%$, $32\%$, $19\%$, and $23\%$ of the sample. 

\begin{table}[h]
\caption{Sample format of patient information collection \label{data_format}}
 \centering
 \resizebox{\columnwidth}{0.9cm}{
\begin{tabular}{ccccccccc} 
 \hline
 \bf Patient ID  & \bf Registration time &  \bf Arrival mode &\bf Age &  \bf Sex & \bf Triage & \bf  Examination time  & \bf Departure time& \bf Destination \\ 
\hline
818897&  2015-01-01 00:54 & Ambulance &   36 &  F &  00:59& 03:00 & 2015-01-01 09:00&Discharged home\\
\hline
818954&  2015-01-01 12:39 & Other  & 78  &  F &   12:54 &      - & 2015-01-01 13:22 & Did not wait\\
\hline
\dots &  \dots  & \dots & \dots	& \dots & \dots & \dots & \dots & \dots\\
\hline
\end{tabular}
}
\end{table}
  
\subsection{ Model fitting and results}
 
\begin{figure}[t]
\centering
 \resizebox {\columnwidth} {!} {
\begin{tikzpicture}
        \tikzset{node style/.style={state, 
                                    fill=white!95!black,
                                    rectangle}}
    
        \node[node style]               (Ia)   {$\theta^1_{1}$};
        \node[node style, right=of Ia]   (IIa)  {$\theta^1_{2}$};
        \node[draw=none,  right=of IIa]   (k-na) {$\cdots$};
        \node[node style, right=of k-na]  (na)   {$\theta^1_{n_1}$};
        
        \tikzset{node style/.style={state, 
                                    fill=white!80!black,
                                    rectangle}}
          \node[node style, right=of na]  (Ib)   {$\theta^2_{1}$};
        \node[node style, right=of Ib]   (IIb)  {$\theta^2_{2}$};
        \node[draw=none,  right=of IIb]   (k-nb) {$\cdots$};
        \node[node style, right=of k-nb]  (nb)   {$\theta^2_{n_2}$};
        
        \tikzset{node style/.style={state, 
                                    fill=white!60!black,
                                    rectangle}}
         \node[node style, right=of nb]  (Ic)   {{$\theta^3_{1}$}};
       \node[node style, right=of Ic]   (IIc)  {$\theta^3_{2}$ };
       \node[draw=none,  right=of IIc]   (k-nc) {$\cdots$};
       \node[node style, right=of k-nc]  (nc)   {$\theta^3_{n_{3}}$};

       \node[draw=none, left=of Ia] (left) {};
       \node[draw=none,  above=of IIa,outer sep=0.1cm]   (station1) {$S_1$};
       \node[draw=none,  above=of IIb,outer sep=0.1cm]   (station2) {$S_2$};
       \node[draw=none,  above=of IIc,outer sep=+0.1cm]   (station3) {$S_3$};
       
        \node[node style,rectangle,rounded corners=10pt,fill=white!60!white,minimum height=1cm,minimum width=3.5cm,node distance=1.2cm, below=of IIa] (qa) {ED exit};
        \node[node style,rectangle,rounded corners=10pt,fill=white!60!white,minimum height=1cm,minimum width=3.5cm,node distance=1.2cm, below=of IIb] (qb) {ED exit};
        
       \node[node style,rectangle,rounded corners=10pt,fill=white!60!white,minimum height=1cm,minimum width=3.5cm,node distance=1.2cm, below=of IIc] (qc) {ED exit};
        
    \draw[>=latex,auto=left,every loop]
         (left)   edge node {}        (Ia) 
           (Ia) edge node {}(IIa)
  (IIa)  edge node {}(k-na)
   (k-na)  edge node {}  (na)
       (Ia)  edge node {$\pi^1_{1}$} (qa)
      (IIa)  edge node {$\pi^1_{2}$}  (qa)
       (na)   edge node {$\pi^1_{n_1}$}   (qa)
        (Ia)  edge [bend left]  node {$\pi^1_{11}$} (Ib)
      (IIa)  edge [bend left]  node {$\pi^1_{21}$}  (Ib)
       (na)   edge [bend left]  node {$\pi^1_{n_11}$}   (Ib)

   (Ib) edge node {}(IIb)
  (IIb)  edge node {}(k-nb)
   (k-nb)  edge node {}  (nb)
       (Ib)  edge node {$\pi^2_{1}$} (qb)
      (IIb)  edge node {$\pi^2_{2}$}  (qb)
       (nb)   edge node {$\pi^2_{n_2}$}   (qb)
       (Ib)  edge [bend left]  node {$\pi^2_{11}$} (Ic)
      (IIb)  edge [bend left]  node {$\pi^2_{21}$}  (Ic)
       (nb)   edge [bend left]  node {$\pi^2_{n_21}$}   (Ic)
       
       (Ic) edge node {}(IIc)
  (IIc)  edge node {}(k-nc)
   (k-nc)  edge node {}  (nc)
       (Ic)  edge node {$\pi^3_{1}$} (qc)
      (IIc)  edge node {$\pi^3_{2}$}  (qc)
       (nc)   edge node {$\pi^3_{n_3}$}   (qc);

\end{tikzpicture}
}   
\caption{ Successive Coxian distributions to model the ED at UHL. \label{Fig_seriesCox2}}
\end{figure}
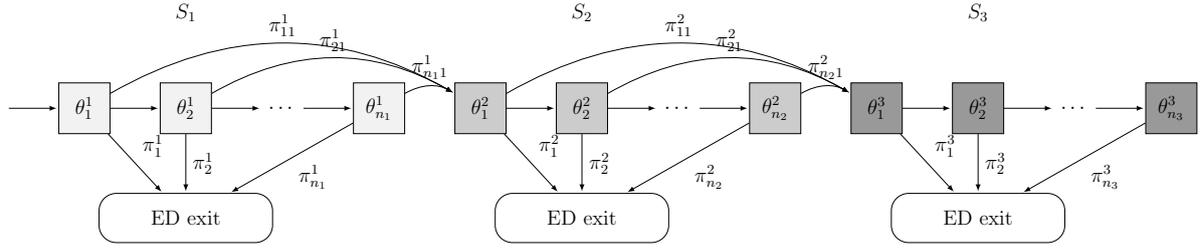

 A conditional Coxian distribution (see Section \ref{section2.7}) was fitted for each of the three stations of the ED; a diagram of the model is given in Figure~\ref{Fig_seriesCox2}. The models were fitted first without including the covariates. The aim of doing so is to investigate the effect of covariates on the goodness of fit and on the number of latent phases in each station. To obtain the number of phases in each station we fitted sequentially an increasing number of phases, starting with one phase, until little improvement in the AIC and BIC values is obtained by adding a new phase. Table~\ref{table3} summarises the results of the fitting process before and after including the covariates. It is clear that the incorporation of the covariates in the model results in smaller AIC and BIC values, and also a reduction in the number of phases. In stations $S_2$ and $S_3$, the number of phases reduced from six to four. This indicates that the heterogeneity in patient length of stay is better explained by covariates than by increasing the number of latent phases. On the other hand, in station $S_1$, the most suitable fit is a five-phase conditional Coxian distribution. The inclusion of covariates does not reduce the number of phases in that station, however, the AIC and BIC values decrease after inclusion the covariates indicating a better fit to the data.

\begin{figure}
\centering
\begin{subfigure}[b]{1\textwidth}
   \includegraphics[width=1\linewidth, height=6.3cm]{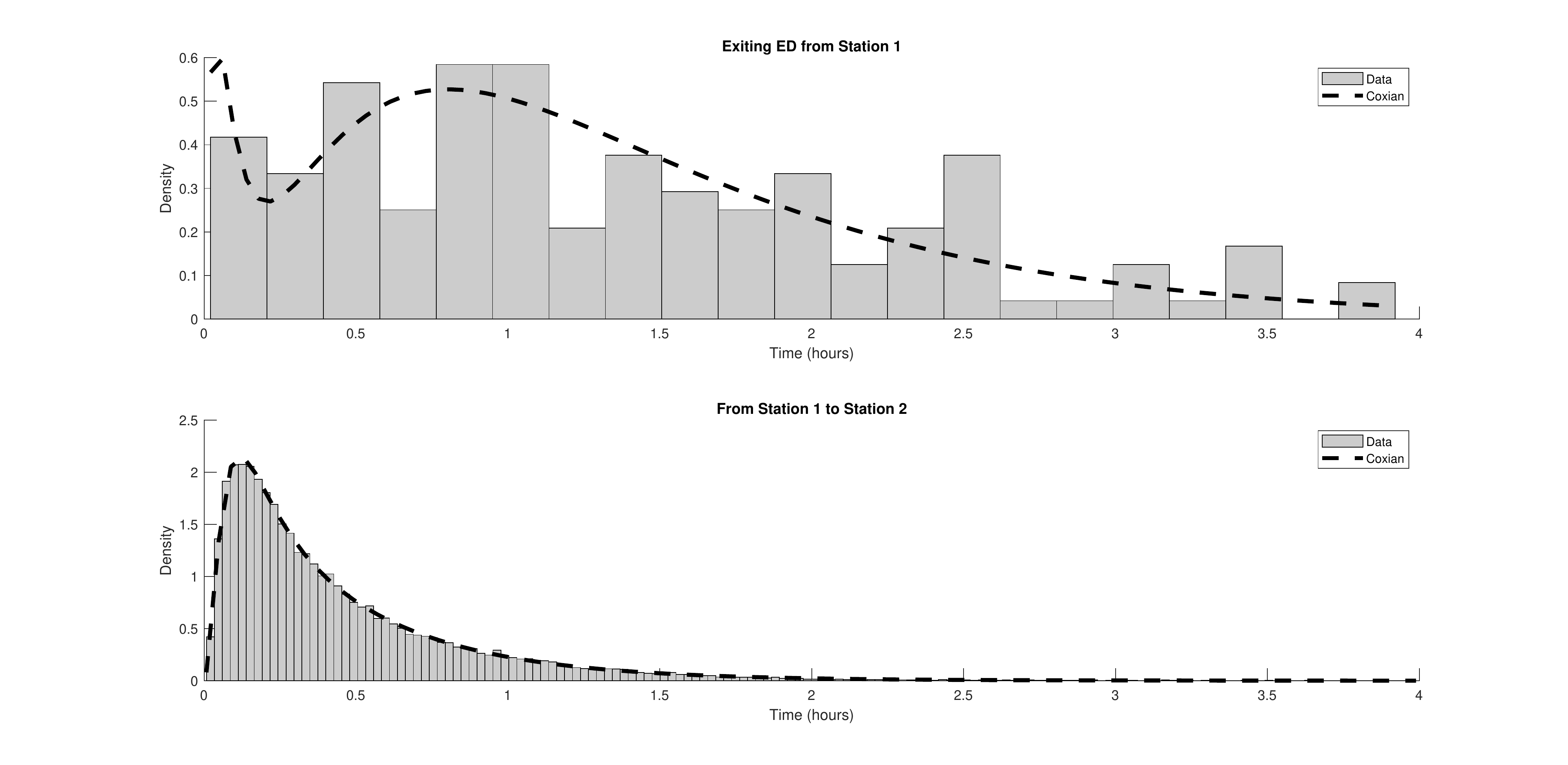}
   \caption{}
   \label{fig:Ng1} 
\end{subfigure}

\begin{subfigure}[b]{1\textwidth}
   \includegraphics[width=1\linewidth, height=6.3cm]{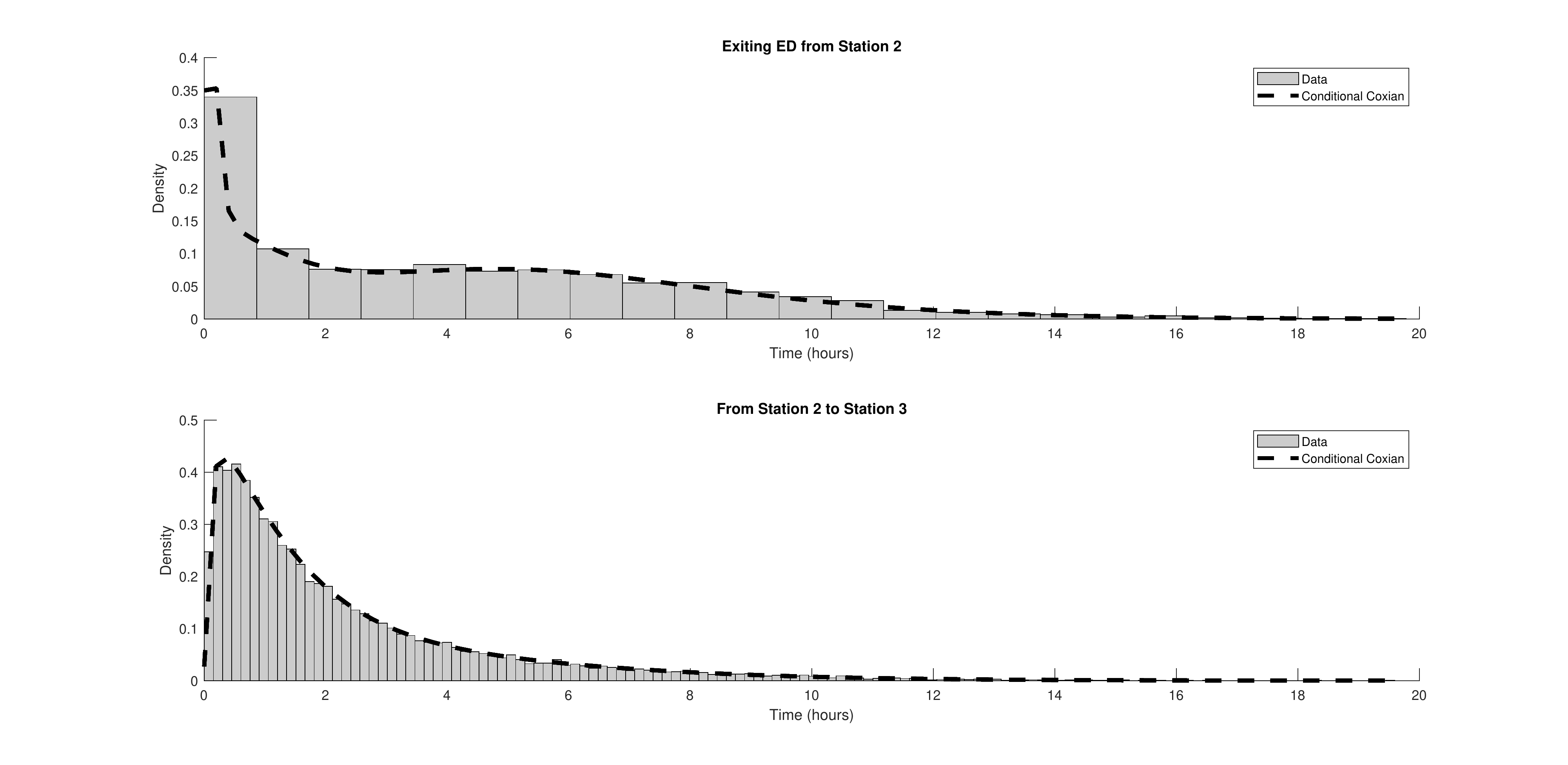}
   \caption{}
   \label{fig:Ng2}
\end{subfigure}
\vspace{-.2cm}
\begin{subfigure}[b]{1\textwidth}
   \includegraphics[width=1\linewidth, height=5.7cm]{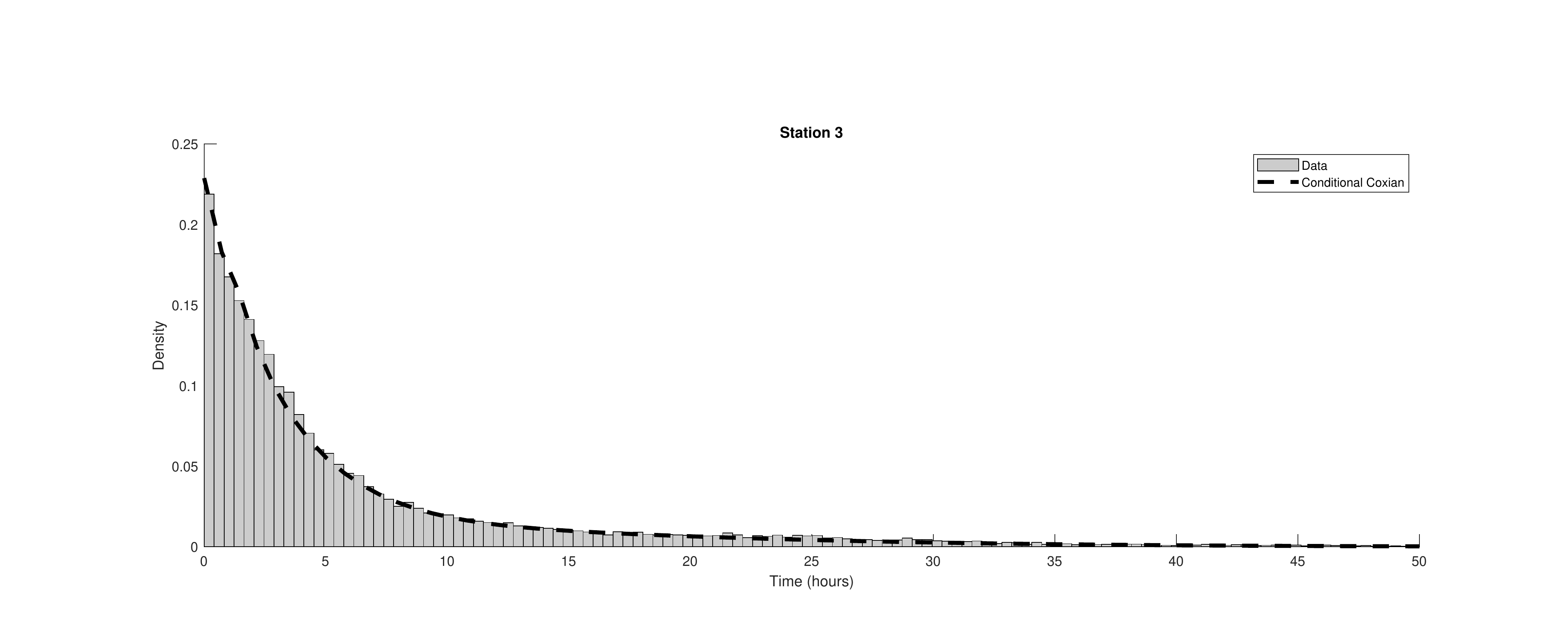}
   \caption{}
   \label{fig:Ng2}
\end{subfigure}

\caption[]{(a) Model fits for the patients who exit the ED from $S_1$ (top) and those who proceed to $S_2$ (bottom). (b) Model fits for the patients who exit the ED from $S_2$ (top) and those who proceed to $S_3$ (bottom). (c) Model fit for the patients receiving treatment in the final station. \label{modelfit} }
\end{figure}

Figure \ref{modelfit} shows the fitted LoS distributions from the optimal model (phases selected using BIC) along with the observed LoS histograms. The fit to the data is excellent in all cases apart from the unusual group who exit ED from $S_1$ (representing only 0.35$\%$ of the sample); nevertheless, the general shape in that case is still reasonably well captured.
 Table \ref{table4}  displays the estimated covariate effects along with their standard errors associated with the optimal model (selected using BIC) fit for each station of the ED.  We see that arrival time, arrival mode, and age play a significant role in patient LoS, whereas the effect of sex is not significant. 
 
On the basis of the estimates, the patients arriving at night spend less time in stations $S_1$ (before triage) and $S_3$ (treatment) than the day patients.  This is due to the fact that the emergency department might be less busy at night time; however, the same  night patients tend to wait more in station $S_2$ (after triage and before treatment), and this is perhaps an indication of lack of available night-shift staff in the treatment area.    For the arrival mode, and as expected, the patients arriving by ambulance are regarded as more urgent, proceeding faster through the first two stations. Indeed, these do tend to be more severe cases, and their mean LoS in $S_3$ (treatment) is significantly longer (approx. 1.7 times) than those who do not arrive by ambulance.   Finally, the age covariate shows a significant effect whereby the youngest group spend less time in $S_2$ and $S_3$, and the older groups spend more time in $S3$; perhaps, older patients have more complex medical needs which necessitate a longer treatment time.

\begin{table}[h]
\caption{Determining the number of phases in each ED station without (null model) and with inclusion of covariates. \label{table3}}
 \centering
 \resizebox{\columnwidth}{1.6cm}{
\begin{tabular}{c|cc|cc||cc|cc||cc|cc} 
 \hline
&\multicolumn{4}{c||}{\bf Registration ($\bf S_1$)}& \multicolumn{4}{c||}{\bf Triage ($\bf S_2$)}&\multicolumn{4}{c}{\bf Treatment ($\bf S_3$)} \\
 \hline
 \bf Phases & \multicolumn{2}{c|}{ Null} &  \multicolumn{2}{c||}{Covariates} &\multicolumn{2}{c|}{Null} &  \multicolumn{2}{c||}{Covariates} & \multicolumn{2}{c|}{Null} & \multicolumn{2}{c}{Covariates} \\ 
 \hline
&    \bf AIC & \bf BIC & \bf AIC &\bf BIC & \bf AIC & \bf BIC &\bf AIC &\bf BIC  & \bf AIC &\bf BIC  &\bf AIC &\bf BIC \\
\hline
1&    16816.25&16841.82& 16287.89&16364.61 &  160738.44 & 160764.25& 158098.91 &158175.60 &203770.51& 203787.40 & 193505.24 &193572.78\\
\hline
2 &  13022.27  &  13073.41&12247.70 & 12349.99& 159742.47	& 159793.60 & 157340.45& 157442.70&199645.04 & 199678.82& 192936.15 & 193020.58\\
\hline
3 &  12583.74 & 12660.46 & 11810.41 &11938.27 &158034.74 	 & 158111.41 &155865.95& 155993.77& 199439.64&199490.30&  192613.66& 192714.97\\
\hline
4 &  12245.69  &12347.98 & 11357.26 & 11510.69&  157727.65	&  157829.93 & \bf 155810.67& \bf 155964.05&199090.76& 199023.22&  \bf 192555.41& \bf 192673.60\\
\hline
5 &  \bf{12147.63}  &\bf{12275.50}& \bf 11273.55  & \bf  11452.56 &157637.65 	&  157765.50& 155816.66 & 155995.60&198949.26& 198864.83& 192543.70&192678.78\\
\hline
6 &  12154.07 &12307.45 & 11253.53 & 11458.11&\bf{157605.58} 	& \bf{ 157759.00} & -& -&\bf 198712.64& \bf 198813.96 &- &-\\
\hline
7 &  -  &-& -  & - &157609.45 	& 157790.16& - & -&198700.00 &198817.42& - &-\\
\hline
\end{tabular}
}

\vspace{1ex}
 {\raggedleft Bold text indicates lowest BIC.
 }

\end{table}

\begin{table}[t]
 \caption{Parameter estimates of the fitted model with covariates, with standard errors in brackets.  \label{table4}}
 \centering
\begin{tabular}{c|c|c||c|c||c|c}
\hline
\multirow{2}{*}{Covariate} & \multicolumn{2}{c||}{ \bf Registration ($\bf S_1$)} &  \multicolumn{2}{c||}{\bf  Triage ($\bf S_2$)}& \multicolumn{2}{c}{ \bf Treatment ($\bf S_3$)}\\
\cline{2-7}
 & $\hat \beta$ & $\exp{(\hat \beta)}$ & $\hat \beta$ & $\exp{(\hat\beta)}$ &$\hat \beta$ & $\exp{(\hat \beta)}$ \\
 \hline
 Arrival time: Night &-0.081$^{**}$  & 0.922& 0.277$^{**}$ & 1.319&-0.119$^{**}$  &0.888 \\
          30$\%$       &(0.015) &      & (0.045) &      & (0.017) &  \\
 \hline
 Arrival mode: Ambulance &-0.312$^{*}$  & 0.732&-0.027$^{*}$ & 0.973&0.512$^{**}$  &1.669 \\
 26$\%$&(0.147) & &(0.012)& &(0.020) & \\
\hline
 Sex: Female  &0.035  &1.036 &0.065 &1.067 &0.046  &1.047 \\
 48$\%$& (0.083) & &(0.034) & & (0.060) & \\
\hline
Age: $<18 $ & -0.0006  &  1.000& -0.462$^{**}$ &0.630 &-0.235$^{**}$ &0.791 \\
 26$\%$&  (0.108) &  &  (0.076)& & (0.044) & \\
   \hline
 Age: $45-64 $ & 0.027 &  1.030& -0.047$^{*}$ & 0.954 &0.444$^{**}$ &1.559 \\
 19$\%$ &  (0.017)& & (0.019)& & (0.038)& \\
 \hline
 Age: $> 65 $ & 0.063  & 1.065 & -0.044 &0.957 & 0.875$^{**}$ & 2.400 \\
23$\%$ & (0.204) & &  (0.040)& & (0.041)&\\
\hline
\end{tabular}

 \vspace{1ex}
 {\raggedleft $*$ and $**$ correspond to 5$\%$ and 1$\%$ significance levels respectively. }
\end{table}

 \newpage
\section{Conclusion}
Having reliable measures for modelling and predicting patient LoS in the emergency department is vital for any hospital. In the existing literature, conditional CPH distributions have been used to model compartmental healthcare systems. However, the assumption of a common LoS distribution for all patients was made (i.e., covariates were absent), and, hence, heterogeneity in patients was not accounted for; this would surely result in producing less accurate predictions, and, furthermore, the insight that covariate effects provide is important. Covariates were not included perhaps due to the somewhat complex structure of the Coxian model, and the consequent estimation instability and large fitting times. 
In this work, we overcome these problems by reformulating the conditional Coxian model using a parametric family of density functions. The new reformulation permits straightforward inclusion of covariates as well as speeding up the fitting process. 

The model has proved useful in the context of the UHL emergency department, where each of three stations (registration, triage, and treatment) was modelled using a reformulated Coxian model. We find that the inclusion of covariates results in a reduced number of phases when fitting the data, and lower AIC and BIC values, i.e., the heterogeneity in patient stay can be better explained using covariates than by increasing the number of phases. This demonstrates the importance of incorporating covariates in CPH models, and, in the context of the UHL ED data, we have found that the arrival time, arrival mode, and patient age are important covariates.  Our proposed model  will assist ED clinicians and managers in understanding the effects of patient characteristics on the demand for resources, and, in particular, assists in identifying groups with particularly long durations.

%
%

\newpage

\bibliographystyle{apa}
\bibliography{references_paper2}

\end{document}